# Dipolar interactions enhanced by two-dimensional dielectric screening in few-layer van der Waals structures

Yuhang Hou[1], Hongyi Yu[1,2*]

[1] Guangdong Provincial Key Laboratory of Quantum Metrology and Sensing & School of Physics and Astronomy, Sun Yat-Sen University (Zhuhai Campus), Zhuhai 519082, China
[2] State Key Laboratory of Optoelectronic Materials and Technologies, Sun Yat-Sen University (Guangzhou Campus), Guangzhou 510275, China
* E-mail: yuhy33@mail.sysu.edu.cn

**Abstract:** We theoretically examined how the dielectric screening of two-dimensional layered materials affects the dipolar interaction between interlayer excitons in few-layer van der Waals structures. Our analysis indicates that the dipolar interaction is largely enhanced by two-dimensional dielectric screening at an inter-exciton separation of several nanometers or larger. The underlying mechanism can be attributed to the induced-charge densities in layered materials, which give rise to induced-dipole densities at large distances with directions parallel to that of the interlayer exciton. The interaction between quadrupolar excitons in trilayer structures are found to be enhanced even larger, with a magnitude one to two orders stronger than that without 2D dielectric screening. The strengths of these dipolar and quadrupolar interactions can be further tuned by engineering the dielectric environment.

## 1. Introduction

Transition metal dichalcogenides (TMDs) have emerged as a fascinating class of two-dimensional (2D) materials, drawing significant attention due to their remarkable electronic and optoelectronic properties [1-3]. The ability of stacking several monolayers into van der Waals few-layer structures has further made them a promising platform for engineering artificial material systems and simulating novel condensed matter phases [4-7]. In recent years, a specific topic of interest has been the interlayer excitons (IXs) in few-layer TMDs, which are novel bound states formed by the Coulomb interaction between an electron and a hole localized in different TMDs layers [8, 9]. Combined with the formation of long-wavelength moiré patterns, these IXs exhibit intriguing properties including the ultralong population and valley lifetimes [10], position-dependent spin and valley optical selection rules [11], localization by the moiré potential [12-14] or inhomogeneous external electric potential [15-17], offering profound implications for next-generation excitonic devices [18].

The spatial separation of the electron and hole in different layers results in a permanent out-of-plane electric dipole for the IX, which plays an important role for tailoring the optical properties of few-layer TMDs structures. Under an out-of-plane electric field, the linear Stark shift proportional to the electric dipole can affect the energy and radiative recombination rate of the IX [9, 19]. Furthermore, the dipolar interaction between IXs leads to a density dependent energy shift which underlies the intriguing diffusion dynamics of IXs [10, 19, 20]. With the presence of a moiré pattern, IXs localized at different moiré potential minima are correlated through their dipolar interaction. The energy of a localized IX then depends on the nearby local occupations which can be shifted by several meV [21, 22]. In a homobilayer moiré pattern, it is suggested that the luminescence properties of localized IXs can be substantially changed by the near-neighbor dipolar forces [23]. Such a strong dipolar interaction can also facilitate the formation of correlated insulators of IXs, as being observed in recent experiments [24-28]. In most of the early literatures, the dipolar interactions are approximated by the traditional form applicable in a 3D homogeneous dielectric medium. However, the dielectric screening of atomically thin 2D materials exhibits a non-local character which is not a constant but varies with the distance between the charges [29-31]. As a result, a non-hydrogenic exciton Rydberg series emerges in monolayer TMDs [32-34]. Currently, a comprehensive understanding to the effect of the 2D dielectric screening on the dipolar interaction between IXs is still lacking. This gap in knowledge could impede the progress in understanding multiparticle configurations of IXs and their correlation phenomena.

Here, we theoretically analyze the dipolar interaction between in-plane separated IXs in bilayer and trilayer TMDs van der Waals structures, taking into account the 2D nature of the dielectric screening from layered materials. We begin in Section 2 by deriving the dipolar interaction form in bilayer TMDs, highlighting its anomalous enhancement by the 2D dielectric screening, and analyzing the underlying mechanism through the induced-charge densities. Then in Section 3, we calculate the various dipolar/quadrupolar interactions in trilayer TMDs, and show that the quadrupolar interaction exhibits a giant enhancement which can be one to two orders of magnitude stronger than that without 2D dielectric screening. In Section 4 we investigate the dielectric engineering of the dipolar interaction strength. Finally, in Section 5 we give a brief summary and discuss the experimental relevance and applicability of our model.

## 2. Dipolar interaction in bilayer TMDs under 2D dielectric screening

We consider a few-layer TMDs/hBN/TMDs system, where two monolayer TMDs are vertically separated by $n$ layers of hexagonal boron nitride (hBN) with $n = 0, 1, 2, \ldots$, see Fig. 1(a) for an illustration. For a low-energy electron and hole located in the top and bottom TMDs layers, respectively, they can form an IX through the long-ranged Coulomb interaction [9]. Such an IX carries a permanent electric dipole $-D$ pointing out-of-plane, with $D = 1.0 + 0.34(n − 1)$ nm the vertical distance between the two TMDs layers. Below we focus on the dipolar interaction between two IX wave packets, by approximating the electron and hole constituents as point charges. Such an approximation is good when the in-plane separation between IXs is much larger than the spatial extension of each IX (determined by the Bohr radius and the wave packet width, typically ~ 2 nm [35-39]). The interaction between IXs is affected by the 2D dielectric screenings of TMDs and hBN layers. Considering their atomically thin geometries, only the in-plane component $\mathbf{E}_\parallel$ of the electric field gets screened by these 2D materials. $\mathbf{E}_\parallel$ induces a finite 2D polarization density $\kappa \mathbf{E}_\parallel$ in a given layer, with $\kappa$ the polarizability of the monolayer material. The dipolar potentials in the top and bottom TMDs layer can then be solved through Poisson's equation [40].

We denote the bottom TMDs and top TMDs layers as layer-1 and layer-$(n+2)$, respectively, whose spatial coordinates along the out-of-plane ($z$) direction are $z_1 = 0$ and $z_{n+2} = D$ (see Fig. 1(a)). The hBN layers are denoted as layer-2, …, layer-$(n+1)$, with $z$-coordinates $z_2$, …, $z_{n+1}$. When a point source-charge is placed at the coordinate origin of the bottom TMDs layer with a density distribution $\delta(\mathbf{r})\delta(z)$, the resultant electric potential $V(\mathbf{r},z)$ satisfies

$$\delta(\mathbf{r})\delta(z) = -\frac{\epsilon}{4\pi}\nabla^2 V(\mathbf{r},z) - \sum_j \kappa_j \nabla_\parallel^2 V(\mathbf{r},z_j)\delta(z-z_j). \tag{1}$$

Here $\mathbf{r} \equiv (x, y)$ is the 2D spatial coordinate, $\nabla \equiv \left(\frac{\partial}{\partial x}, \frac{\partial}{\partial y}, \frac{\partial}{\partial z}\right)$ and $\nabla_\parallel \equiv \left(\frac{\partial}{\partial x}, \frac{\partial}{\partial y}\right)$. On the above right-hand-side, the first term is the contribution from the dielectric background with a dielectric constant $\epsilon$, and the second term accounts for the total screening effect of all TMDs and hBN layers with $\kappa_j$ the polarizability of the $j$-th layer. We set $\kappa_1 = \kappa_{n+2} = \kappa_{\mathrm{TMD}} \approx 0.7$ nm [30, 31] and $\kappa_2 = \cdots = \kappa_{n+1} = \kappa_{\mathrm{hBN}} \approx 0.1$ nm [41]. By Fourier transforming to the momentum space, $V(\mathbf{q},z_j) = \int V(\mathbf{r},z_j)e^{-i\mathbf{q}\cdot\mathbf{r}}d\mathbf{r}$ can be solved for all layer index $j$ from Eq. (1).

Here we only consider the intralayer Coulomb potential $V_{\mathrm{intra}}(q) \equiv V(q,0)$ and interlayer potential $V_{\mathrm{inter}}(q) \equiv V(q,D)$ in the two TMDs layers, as shown in Fig. 1(a). Taking $n = 1$ as example, $V_{\mathrm{intra}}$ and $V_{\mathrm{inter}}$ can be expressed as

$$V_{\mathrm{intra}}(q) = \frac{2\pi[1 + qr_0 - qr_0 e^{-2Dq} + qr_0'(1 - e^{-Dq})(1 + qr_0 - qr_0 e^{-Dq})]}{\epsilon q(1 + qr_0 - qr_0 e^{-Dq})[1 + qr_0 + qr_0 e^{-Dq} + qr_0'(1 + qr_0 - qr_0 e^{-Dq})]},$$

$$V_{\mathrm{inter}}(q) = \frac{2\pi e^{-Dq}}{\epsilon q(1 + qr_0 - qr_0 e^{-Dq})[1 + qr_0 + qr_0 e^{-Dq} + qr_0'(1 + qr_0 - qr_0 e^{-Dq})]}. \tag{2}$$

Here $r_0 = 2\pi\kappa_{\mathrm{TMD}}/\epsilon$ and $r_0' = 2\pi\kappa_{\mathrm{hBN}}/\epsilon$ are 2D screening lengths of monolayer TMDs and hBN, respectively. For theoretical insights, below we treat $r_0$, $r_0'$ and $\epsilon$ as independent parameters unless specified. Note that by setting $r_0' = 0$, the intralayer and interlayer Coulomb interaction forms become exactly the same as those for bilayer TMDs without hBN [40]. The real-space forms $V_{\mathrm{intra/inter}}(r) \equiv (2\pi)^{-2}\int d\mathbf{q} V_{\mathrm{intra/inter}}(q)e^{i\mathbf{q}\cdot\mathbf{r}}$ can then be obtained from Eq. (2), which are shown in Fig. 1(b) inset for three different $r_0$ values. Due to the 2D dielectric screening, both $V_{\mathrm{inter}}$ and $V_{\mathrm{intra}}$ weaken with the increase of $r_0$. The dipolar interaction $V_D^{(B)}(r) = 2V_{\mathrm{intra}}(r) - 2V_{\mathrm{inter}}(r)$ has the following form

$$V_D^{(B)}(r) = \frac{2}{\epsilon}\int_0^\infty \frac{(1 - e^{-Dq})J_0(qr)}{1 + qr_0(1 - e^{-Dq})}dq. \tag{3}$$

Here $J_0$ is the Bessel function of the first kind. Interestingly, the parameter $r_0'$, representing the screening length of the hBN material, does not appear in this equation. Such a conclusion for $V_D^{(B)}$ only applies to $n = 1$ but not for $n \geq 2$ (see Appendix for a detailed analysis). Nevertheless, for $n = 2$, 3 and 4 layers of hBN we have calculated the corresponding dipolar interaction in Appendix, and found that the effect of $r_0'$ can still be neglected. The insertion of hBN layers then mainly serves to tune $D$. Also note that for $r_0 = 0$, $V_D^{(B)}(r)\big|_{r_0=0} = V_D^{(3D)}(r) \equiv \frac{2}{\epsilon}\left(\frac{1}{r} - \frac{1}{\sqrt{r^2+D^2}}\right)$ returns to the traditional dipolar interaction form in a 3D homogeneous dielectric medium. When $r \gg D$, Eq. (3) can be approximated as $V_D^{(B)}(r)\big|_{r\gg D} \approx \frac{2}{\epsilon r_0} K_0\left(\frac{r}{\sqrt{r_0 D}}\right) + \frac{D^2}{\epsilon r^3}$ with $K_0$ the modified Bessel function of the second kind.

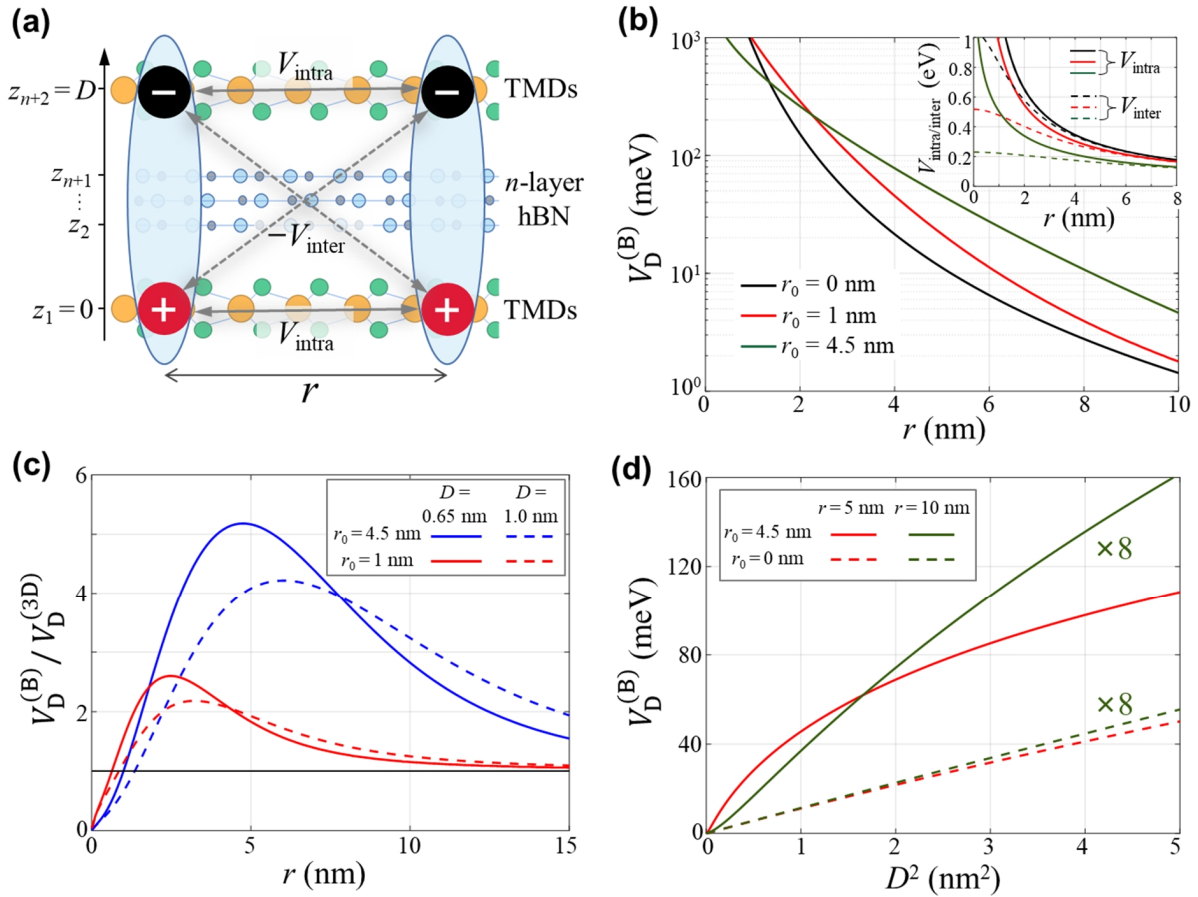

Figure 1. (a) A schematic illustration of the few-layer TMDs/hBN/TMDs system, which consists of $n$ layers of monolayer hBN sandwiched between two monolayer TMDs. Two IXs with electrons in the top TMDs layer and holes in the bottom TMDs layer interact through the intralayer and interlayer Coulomb potentials. (b) The dipolar interaction strength $V_D^{(B)}$ as a function of the in-plane separation $r$ for $D = 1.0$ nm and $\epsilon = 1$. (c) The enhancement factor $V_D^{(B)}/V_D^{(3D)}$ as a function of $r$. (d) $V_D^{(B)}$ as a function of $D^2$. For clarity, the curves of $r = 10$ nm are multiplied by a factor of 8.

Fig. 1(b) shows our calculated values of $V_D^{(B)}(r)$ as functions of $r$. With the increase of $r_0$, the behavior of $V_D^{(B)}(r)$ is very different from the Coulomb interaction $V_{\text{intra/inter}}$ shown in Fig. 1(b) inset. Near $\mathbf{r} = 0$, $V_D^{(B)}(r)$ is weakened by the 2D dielectric screening of TMDs layers just as expected. However, for $r$ in the order of several nm or larger which is the typical separation between nearest-neighbor moiré potential minima, $V_D^{(B)}(r)$ shows an anomalous enhancement by the 2D screening length $r_0$, which has also been notices in several early works [42, 43]. This can be seen more clearly from the calculated enhancement factor $V_D^{(B)}/V_D^{(3D)}$ shown in Fig. 1(c), which indicates that $V_D^{(B)}$ can be several times larger than $V_D^{(3D)}$ at a distance of several nm. The maximum enhancement factor occurs at a critical distance **~2.54** $\sqrt{r_0 D}$, with $\max\left(V_D^{(B)}/V_D^{(3D)}\right) \propto \sqrt{r_0/D}$. Above this critical distance, $V_D^{(B)}/V_D^{(3D)}$ decreases with $r$ and approach 1 for $r \gg \sqrt{r_0 D}$. As the dipolar interaction gets stronger when increasing $D$ which can be realized by inserting more hBN layers, we show the dependence of $V_D^{(B)}$ on $D^2$ in Fig. 1(d). For experimentally accessible electric dipole values (0.6 nm $< D \sim 1$ nm), $V_D^{(B)}$ shows a steeper slope with $D^2$ than $V_D^{(3D)} \approx \frac{D^2}{\epsilon r^3}$, implying that $V_D^{(B)}$ increases faster with $D$ compared to $V_D^{(3D)}$.

Now we investigate the underlying mechanism of this anomalous enhancement to the dipolar interaction. The difference between $V_D^{(B)}$ and $V_D^{(3D)}$ originates from the 2D induced-charge densities in TMDs layers. For illustration purpose, we consider a positive source-charge with a 2D gaussian profile $\rho_s(r)\delta(z) = \frac{1}{2\pi\sigma^2} e^{-\frac{r^2}{2\sigma^2}}\delta(z)$ centered at the coordinate origin of the bottom TMDs layer, which introduces an electric potential $V_g(\mathbf{r}, z) \equiv \frac{1}{2\pi\sigma^2}\int d\mathbf{r}' V(|\mathbf{r} - \mathbf{r}'|, z) e^{-\frac{r'^2}{2\sigma^2}}$. The induced-charge densities in TMDs layers can be obtained as

$$\rho_{\text{ind}}(r) = \kappa_{\text{TMD}} \nabla_\parallel^2 V_g(\mathbf{r}, 0),$$
$$\rho'_{\text{ind}}(r) = \kappa_{\text{TMD}} \nabla_\parallel^2 V_g(\mathbf{r}, D).$$
(4)

The calculated $\rho_{\text{ind}}(r)$, which corresponds to the induced-charge density in the bottom TMDs layer, and $\rho'_{\text{ind}}(r)$, which is that in the top TMDs layer, are schematically shown in Fig. 2(a). Unlike the induced-charge $-(1 - \epsilon^{-1})\rho_s(r)\delta(z)$ from the 3D homogeneous dielectric medium which is always negative and located at the same position as the source-charge, $\rho_{\text{ind}}(r)$ and $\rho'_{\text{ind}}(r)$ have much wider spatial distributions and both cross from negative at short distances to positive at large distances. In the bottom TMDs layer where the source-charge resides, the induced-charge density $\rho_{\text{ind}}(r)$ shows a negative gaussian shape within a circular region centered at $\mathbf{r} = 0$ (blue-colored region in Fig. 2(a)). $\rho_{\text{ind}}(r)$ crosses zero at a critical radius $r_c$, and becomes positive for $r > r_c$ (red-colored region in Fig. 2(a)). The critical value $r_c$ depends sensitively on the source-charge profile, which decreases with the narrowing of the gaussian width $\sigma$. When the source-charge shrinks to a point charge with $\sigma = 0$, the induced-charge density $\rho_{ind}(r)$ becomes the superposition of $-\delta(\mathbf{r})$ and a positive function with a finite distribution width (see Figs. 2(b)). The origin of these two components can be understood as follows: the intralayer Coulomb potential diverges logarithmically for $r \to 0$ with an asymptotic form $V_{intra}(r)|_{r \to 0} \to \frac{-1}{\epsilon r_0} \ln\left(\frac{r}{2r_0}\right)$, which gives rise to the $\delta$-function component $\kappa_{TMD} \nabla_\parallel^2 V_{intra}(r)|_{r \to 0} = -\delta(\mathbf{r})$; meanwhile, the positive component with a finite distribution width comes from the fact that $\nabla_\parallel^2 V_{intra}(r)|_{r>0} \neq 0$. Specifically, $V_{intra}(r)|_{r \gg r_0} \to \frac{1}{\epsilon r}$ at large distances, where the 2D induced-charge $\kappa_{TMDs} \nabla_\parallel^2 V_{intra}(r)|_{r \gg r_0} \to \frac{r_0}{2\pi r^3} > 0$ has the same sign as the source-charge. A comparison between Coulomb potentials under 3D and 2D dielectric screening effects and the corresponding induced-charge densities are given in Table I, showing distinct behaviors. For the finite but weak induced-charge density $\rho'_{ind}(r)$ in the top TMDs layer, it has a weak dependence on the source-charge width $\sigma$ due to the non-divergence of $V_{inter}(r)$ at $\mathbf{r} = \mathbf{0}$, see Figs. 2(c).

Table I. The comparison between 3D and 2D dielectric screening effects on the Coulomb potential from a positive point charge located at $\mathbf{r} = 0$. $\epsilon$ is the dielectric constant of the 3D homogeneous medium, and $r_0$ is the screening length of the 2D material. The 2D screened Coulomb potential has a complicated form, thus only asymptotic forms at $r \to \mathbf{0}$ and $r \gg r_0$ are given.

|  | 3D case, $\mathbf{r} = (x, y, z)$ |  | 2D case, $\mathbf{r} = (x, y)$ |  |
| --- | --- | --- | --- | --- |
|  | Coulomb potential | Induced-charge density | Coulomb potential (asymptotic form) | Induced-charge density |
| $r \to \mathbf{0}$ | $(\epsilon r)^{-1}$ | $-(1 - \epsilon^{-1})\delta(\mathbf{r})$ | $\frac{-1}{\epsilon r_0}\ln\left(\frac{r}{2r_0}\right)$ | $-\delta(\mathbf{r})$ |
| $r \gg r_0$ | $(\epsilon r)^{-1}$ | 0 | $(\epsilon r)^{-1}$ | $\frac{1}{2\pi} r_0 r^{-3}$ |

A pair of opposite source-charges located in different TMDs layers forms a source electric dipole, see Figs. 2(d). The corresponding induced-charge densities in the bottom and top TMDs layers are given by $\rho_{sum} \equiv \rho_{ind} - \rho'_{ind}$ and $-\rho_{sum}$, respectively. This then gives rise to an induced-dipole density $-\rho_{sum} D$ pointing along $\pm z$, which is finite even far away from the source-dipole. $\rho_{sum}$ varies from negative to positive with position. However, the overall charge density $\rho_s + \rho_{sum}$, including both the source-charge and induced-charge contributions, is positive on the whole 2D plane, see Figs. 2(d). Figs. 2(e) shows the spatial profiles of $\rho_s$ and

$\rho_s + \rho_{\text{sum}}$. As can be seen, the effect of the 2D dielectric screening is to transform the narrowly distributed source-charge $\rho_s$ into a spatially extended overall charge distribution $\rho_s + \rho_{\text{sum}}$, at the cost of a largely reduced maximum density at $\mathbf{r} = 0$. Compared to the fast gaussian decay of $\rho_s$, $\rho_s + \rho_{\text{sum}}$ exhibits a power-law decaying tail ($\propto r^{-5}$) independent on $\sigma$ at large distances. The resultant electric dipole density $-(\rho_s + \rho_{\text{sum}})D$, which is parallel to the source-dipole, can then phenomenologically explain how the 2D dielectric screening enhances the dipolar interaction while keeps weakening Coulomb interactions. At a position of several nm away from the source-charge, the long-ranged Coulomb interaction is dominated by the overall charge density near $\mathbf{r} = 0$, which is largely reduced by the 2D dielectric screening (Figs. 2(e)). On the other hand, the dipolar interaction is much shorter-ranged than the Coulomb interaction, it is thus dominated by the induced-dipole density at this position, where the slowly decaying tail of the dipole density enhances $V_D^{(B)}$. We also note that the 2D dielectric screening doesn't change the total charge, i.e., $\int_0^\infty dr r \rho_{\text{ind}}(r) = \int_0^\infty dr r \rho'_{\text{ind}}(r) = 0$ (Figs. 2(f)). This is again distinct to that in a 3D homogeneous dielectric medium, where the total charge is reduced by a factor $\epsilon^{-1}$ (see Table I).

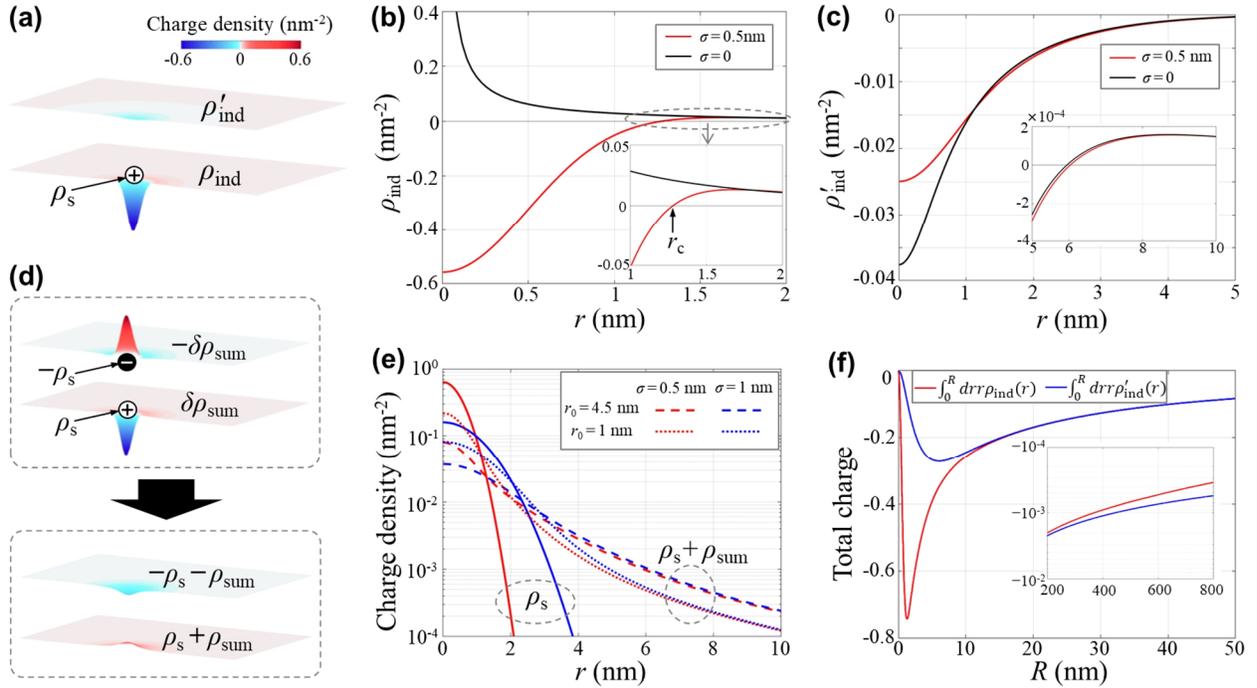

Figure 2. (a) The induced-charge densities $\rho_{\text{ind}}$ and $\rho'_{\text{ind}}$ in the bottom and top TMDs layers, respectively. $\rho_s$ denotes the source-charge in a 2D Gaussian profile with a width $\sigma = 0.5$ nm. (b) The radial distribution of $\rho_{\text{ind}}$ in the bottom layer. The red and black lines correspond to $\sigma = 0.5$ nm and $\sigma = 0$, respectively. The inset shows the crossover from negative to positive at a critical distance $r_c$ for $\sigma = 0.5$ nm. $\rho_{\text{ind}}$ under $\sigma = 0$ contains a component $-\delta(\mathbf{r})$ which is not shown. (c) The radial distributions of $\rho'_{\text{ind}}$ in the top TMDs layer. The inset shows that $\rho'_{\text{ind}}$ also crosses from negative to positive at a critical distance. (d) The upper panel shows induced-charge densities $\pm \rho_{\text{sum}}$ from a pair of opposite source-charges $\pm \rho_s$. The lower panel shows the total charge densities $\pm(\rho_s + \rho_{\text{sum}})$. (e) The radial distributions of the source-charge $\rho_s$ (solid lines) and the overall charge $\rho_s + \rho_{\text{sum}}$ (dashed and dotted lines). Red and blue colors correspond to $\sigma = 0.5$ and 1 nm, respectively. Dashed and dotted lines correspond to $r_0 = 4.5$ and 1 nm, respectively. $\rho_s + \rho_{\text{sum}}$ exhibits a power-law decaying tail at large distances. (f)

The total induced-charges $\int_0^R dr r \rho_{ind}(r)$ and $\int_0^R dr r \rho'_{ind}(r)$ in a circular region with radius $R$. Both $\to 0$ when $R \to \infty$. The parameters are set as $D = 0.65$nm, $r_0 = 4.5$ nm, $\epsilon = 1$, unless specified.

## 3. Dipolar and quadrupolar interactions in trilayer TMDs

It has been suggested that mirror-symmetric trilayer structures of TMDs (e.g., WSe$_2$/MoSe$_2$/WSe$_2$) can serve as a novel platform for exploring correlated phenomena and quantum phase transitions of bosonic particles, due to the field-tunable formation of dipolar and quadrupolar excitons with strong mutual interactions [44]. In such a trilayer structure, the low-energy electron is located in the middle layer whereas the low-energy hole is in the top or bottom layer, or vice versa. The formed IX can have an electric dipole pointing upward or downward, or can be in their coherent superposition which has zero electric dipole but a finite electric quadrupole, see illustrations in Fig. 3(a). Such quadrupolar excitons have been realized and detected in a series of recent experiments [25, 45-47]. In this section, we investigate the effects of 2D dielectric screening on the related dipolar and quadrupolar interactions in trilayer TMDs.

We first consider two dipolar excitons with parallel electric dipoles, with the hole and electron constituents located in the top and middle TMDs layers, respectively (Fig. 3(a)). Compared to the dipolar interaction $V_D^{(B)}$ in a bilayer TMDs (Fig. 1), the additional layer results in a slightly different dipolar interaction $V_D^{(T)}$ in the trilayer TMDs, as can be seen from the calculated enhancement factor $V_D^{(T)}/V_D^{(3D)}$ shown in Fig. 3(b). Although the maximum values of the enhancement factors are nearly the same after increasing the layer number, but the decay speed with $r$ at large distances slows down significantly for more layers. Such an effect is further corroborated by adding a fourth TMDs layer on top of the IX, see Fig. 3(b). This can be understood as the influence of induced-charges in the additional TMDs layers, which increase the strength of the dipolar potential at large distances.

Next, we consider two dipolar excitons with antiparallel electric dipoles, see the illustration in Fig. 3(a). Their interaction $V_{antiD}^{(T)}$ has the following momentum-space form:

$$V_{antiD}^{(T)}(q) = \frac{2\pi(1-e^{-Dq})^2(1+qr_0+qr_0 e^{-Dq})^2}{\epsilon q[(1+qr_0)e^{-2Dq}-qr_0][(1+qr_0)^2+qr_0(1-qr_0)e^{-2Dq}]}. \quad (5)$$

When $r_0 = 0$, the real-space form $V_{antiD}^{(T)}(r)\big|_{r_0=0} = V_{antiD}^{(3D)}(r) \equiv \frac{1}{\epsilon}\left(\frac{1}{r}+\frac{1}{\sqrt{4D^2+r^2}}-\frac{2}{\sqrt{D^2+r^2}}\right)$ returns to the traditional form in a 3D homogeneous dielectric medium. The calculated $V_{antiD}^{(T)}$ as functions of $r$ for different $r_0$ values are shown in Fig. 3(c). All curves show non-monotonic behaviors, which cross from repulsive ($> 0$) at short distances to attractive ($< 0$) at large distances. With the increase of $r_0$, the potential curve flattens such that the strongest attractive potential value becomes more and more close to zero. Meanwhile, the critical distance $r_c$ at which $V_{antiD}^{(T)}(r_c) = 0$ becomes larger. At a large enough $r$, the strength of $V_{antiD}^{(T)}$ is again

enhanced by the 2D dielectric screening, with the enhancement factor $V_{antiD}^{(T)}/V_{antiD}^{(3D)}$ shown in Fig. 3(d). The magnitude of $V_{antiD}^{(T)}/V_{antiD}^{(3D)}$, however, is significantly smaller than $V_D^{(B)}/V_D^{(3D)}$ between parallel electric dipoles shown in Fig. 3(b).

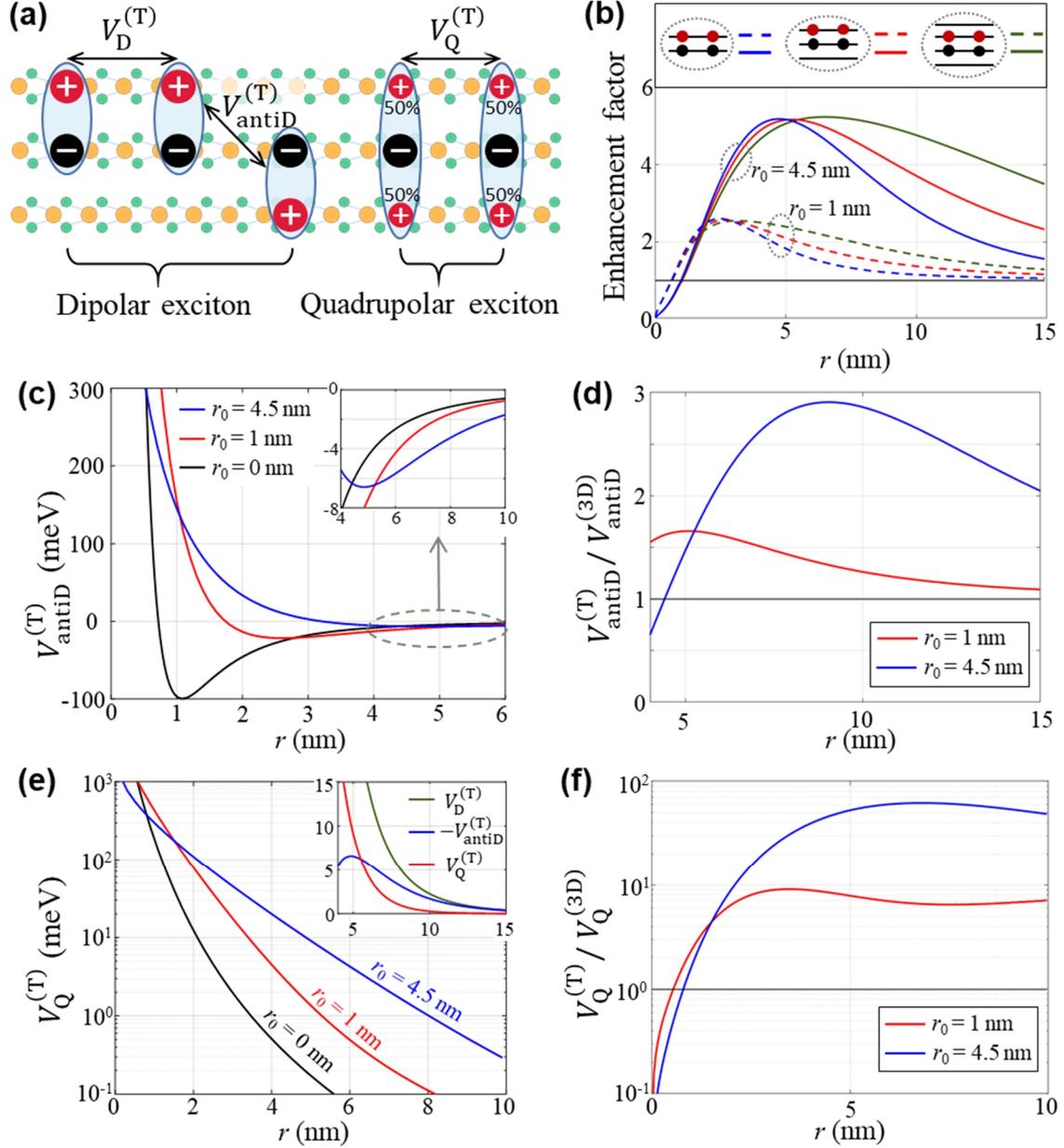

Fig.3 (a) Schematic illustrations of dipolar and quadrupolar excitons in trilayer TMDs, and their mutual interactions. (b) Enhancement factors of dipolar interactions between IXs with parallel electric dipoles, in bilayer (blue curves), trilayer (red curves) and four-layer (green curves) TMDs. (c) Dipolar interaction $V_{antiD}^{(T)}$ between IXs with antiparallel electric dipoles under $\epsilon = 1$. The black, red and blue lines correspond to $r_0 = 0$, 1 and 4.5 nm, respectively. The inset is the enlarged view of $V_{antiD}^{(T)}$ at large distances. (d) The enhancement factor $V_{antiD}^{(T)}/V_{antiD}^{(3D)}$ at large distances. (e) The quadrupolar interaction $V_Q^{(T)}$ between quadrupolar excitons. The inset shows the comparison between $V_D^{(T)}$, $-V_{antiD}^{(T)}$ and $V_Q^{(T)}$ under $r_0 = 4.5$ nm and $\epsilon = 1$. (f) The enhancement factor $V_Q^{(T)}/V_Q^{(3D)}$, which is giant (~ 10-100) in a large range of $r$.

A quadrupolar exciton can form which corresponds to the coherent superposition of the two dipolar excitons with opposite electric dipoles, see Fig. 3(a). The quadrupolar interaction between them has the following momentum-space form:

$$V_Q^{(T)}(q) = \frac{1}{2}V_D^{(T)}(q) + \frac{1}{2}V_{\text{antiD}}^{(T)}(q) = \frac{\pi(1-e^{-Dq})[3+3qr_0+(3qr_0-1)e^{-Dq}]}{\epsilon q[(1+qr_0)^2+qr_0(1-qr_0)e^{-2Dq}]}. \quad (6)$$

Similarly, by setting $r_0 = 0$ we get the traditional quadrupolar interaction $V_Q^{(T)}(r)\big|_{r_0=0} = V_Q^{(3D)}(r) \equiv \frac{1}{2\epsilon}\left(\frac{3}{r} + \frac{1}{\sqrt{4D^2+r^2}} - \frac{4}{\sqrt{D^2+r^2}}\right)$ in a 3D homogeneous dielectric medium. Their interaction strengths for three different values of $r_0$ are shown in Fig. 3(e). Compared to $V_Q^{(3D)}$, the 2D dielectric screening introduces a giant enhancement to $V_Q^{(T)}$, such that its strength can be comparable to the dipolar interaction strength at an experimentally accessible distance $r \sim 5$ nm (Fig. 3(e) inset). The enhancement factor $V_Q^{(T)}/V_Q^{(3D)}$, shown in Fig. 3(f), is found to exhibit a value that can reach $\sim 60$ ($\sim 10$) in a large range of $r$ under $r_0 = 4.5$ nm ($r_0 = 1$ nm).

## 4. Dielectric engineering of the dipolar interaction strength

In 2D layered materials, the Coulomb interaction between charged carriers depends sensitively on the dielectric environment. Experiments have shown that by placing a few-layer graphene (FLG) sheet on top of TMDs, the Coulomb interaction in TMDs can be tuned in an in-situ way by varying the doped carrier density in graphene [48, 49]. In this section we give an analysis on how the dipolar interaction in bilayer TMDs is screened by free carriers in adjacent FLG sheets. We consider a structure shown in Fig. 4(a), where two FLG sheets are placed at the top and bottom sides of the bilayer TMDs, respectively. We treat each FLG sheet as a 2D metal located at a vertical distance $D'$ above/below the bilayer TMDs. In response to the dipolar potential from an IX in TMDs, free carriers in FLG with an average 2D density $\rho_G$ can rearrange their positions and give rise to a density fluctuation $\Delta\rho_G(r)$, which partly neutralizes the source-charges in TMDs as illustrated in Fig. 4(a). Based on the 2D Thomas-Fermi model, the spatial fluctuation of the free carrier density in $l$-th FLG sheet can be written in a form $\Delta\rho_G(r) = -\frac{d\rho_G}{d\mu}V(r,z_l) = -\frac{\epsilon}{2\pi}q_{\text{TF}}V(r,z_l)$, with $z_l$ the $z$-coordinate of the graphene layer, $\mu$ the chemical potential and $q_{\text{TF}} \equiv \frac{2\pi}{\epsilon}\frac{d\rho_G}{d\mu}$ the Thomas-Fermi wavevector. After taking into account the effect of $\Delta\rho_G(r)$, the Poisson's equation for the electrostatic potential $V(r,z)$ induced by a source charge $\rho_s(r,z)$ is now written as

$$\rho_s(r,z) = -\frac{\epsilon}{4\pi}\nabla^2 V(r,z) - \sum_j \kappa_j \nabla_\parallel^2 V(r,z_j)\delta(z-z_j) + \frac{\epsilon}{2\pi}\sum_l q_{\text{TF}}V(r,z_l)\delta(z-z_l). \quad (7)$$

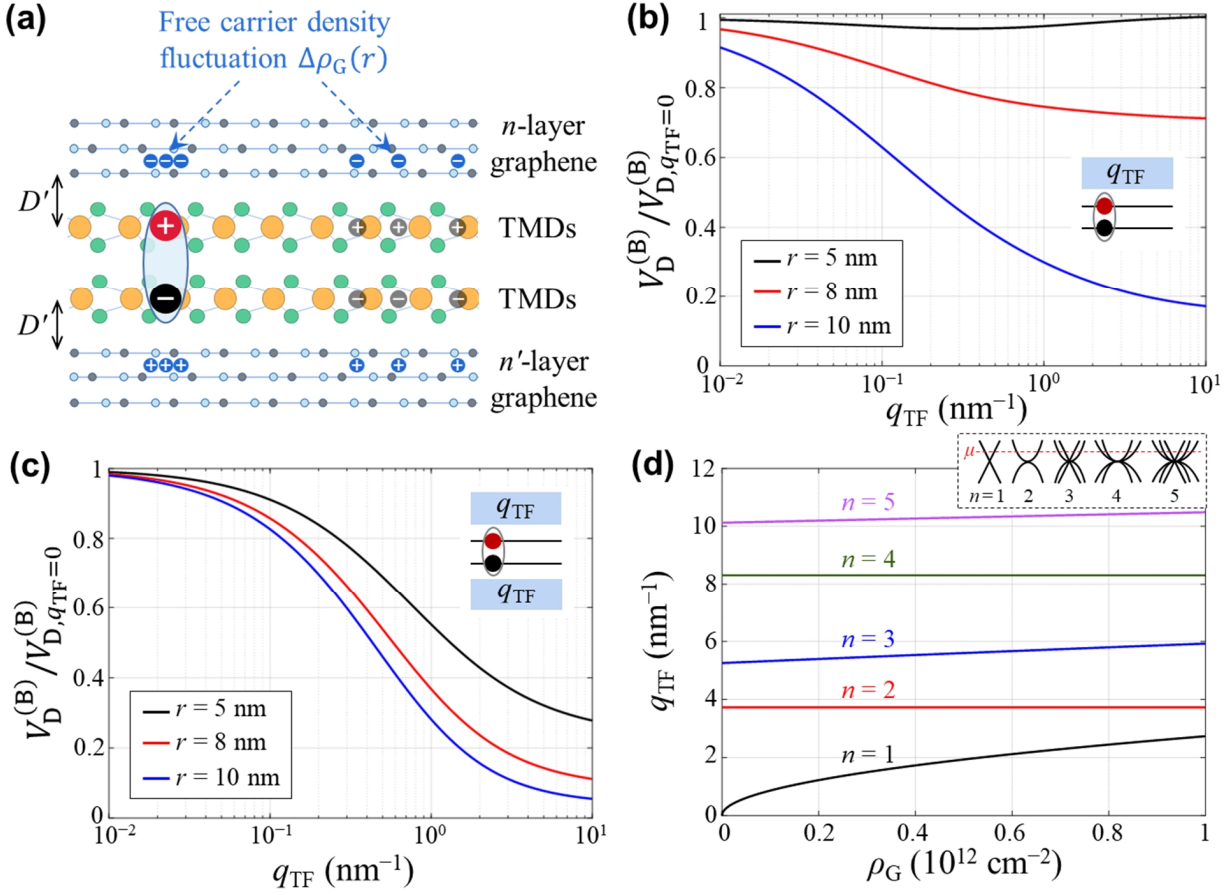

Figure 4. (a) A schematic illustration of the dielectric environment for bilayer TMDs, where an $n$-layer ($n'$-layer) graphene sheet is placed at the top (bottom) side. The dipolar potential from an IX in TMDs leads to free carrier density fluctuations in the two graphene layers. (b) When the graphene sheet is absent at one side of TMDs, the ratio $V_D^{(B)}/V_{D,q_{TF}=0}^{(B)}$ as a function of the Thomas-Fermi wavevector $q_{TF}$ of the remaining graphene sheet. Black, red and blue curves correspond to in-plane distances $r = 5$, 8 and 10 nm, respectively. (c) $V_D^{(B)}/V_{D,q_{TF}=0}^{(B)}$ as a function of $q_{TF}$, when the bilayer TMDs is encapsulated by two graphene sheets. (d) The zero-temperature value of $q_{TF}$ as a function of the carrier density $\rho_G$ in an $n$-layer graphene ($n = 1$ to 5). The inset illustrates low-energy band dispersions for $n = 1$ to 5. $\mu$ is the chemical potential.

We solve the dipolar interaction strength $V_D^{(B)}$ from Eq. (7) for $D' = 0.3$ nm and $\epsilon = 1$. Fig. 4(b,c) show the calculated results of $V_D^{(B)}/V_{D,q_{TF}=0}^{(B)}$ as functions of $q_{TF}$ at $r = 5$, 8 and 10 nm, where $V_{D,q_{TF}=0}^{(B)}$ denotes the dipolar interaction strength at $q_{TF} = 0$. Fig. 4(b) corresponds to the case that the FLG sheet is placed only at one side of TMDs. As can be seen, the strength of $V_D^{(B)}$ is susceptible to the Thomas-Fermi screening of the FLG sheet only at a large enough distance. For a relatively short distance of $r = 5$ nm, $V_D^{(B)}$ is barely affected by $q_{TF}$, indicating that the dipolar interaction becomes much shorter-ranged. When the TMDs layer is encapsulated by two

FLG sheets, the corresponding Thomas-Fermi screening becomes more efficient. The calculated $V_D^{(B)}/V_{D,q_{TF}=0}^{(B)}$ are shown in Fig. 4(c), where we have assumed that the top and bottom FLG sheets have the same $q_{TF}$ value. The three curves with $r = 5$, 8 and 10 nm all decrease significantly with $q_{TF}$, indicating that $V_D^{(B)}(r)$ for a large range of $r$ can be efficiently tuned by screenings of top and bottom FLG sheets. The value of $q_{TF}$ is determined by $\frac{d\rho_G}{d\mu}$ which under a zero temperature corresponds to the density-of-states at the chemical potential. The inset of Fig. 4(d) schematically shows the low-energy band structures of the $n$-layer graphene for $n = 1$ to 5, which contain massless and massive branches. We use a Fermi velocity $v_F = 10^6$ m/s for massless branches of $n = 1$, 3 and 5, and set effective masses of massive branches as $0.039m_0$ ($n = 2$), $0.055m_0$ ($n = 3$), $0.024m_0$ and $0.063m_0$ ($n = 4$), $0.039m_0$ and $0.067m_0$ ($n = 5$) following Ref. [50], with $m_0$ the free electron mass. From the obtained density-of-states of FLG, the resultant zero-temperature $q_{TF}$ values as functions of $\rho_G$ are shown in Fig. 4(d). We can see that $q_{TF}$ can be largely tuned from 0 to 10 nm$^{-1}$ by $\rho_G$ and $n$ for typical carrier densities of $\rho_G < 10^{12}$ cm$^{-2}$. This suggests that the dipolar interaction in few-layer TMDs can be effectively engineered through tuning the doping densities in surrounding graphene layers.

## 5. Summary and discussion

In summary, we have investigated the dipolar interaction (also including the quadrupolar interaction) between IXs in few-layer TMDs structures. Compared to the intralayer and interlayer Coulomb interactions between carriers which are always weakened by the 2D dielectric screening, the dipolar (quadrupolar) interaction is found to be anomalously enhanced by several (several tens) times at an inter-exciton distance of several nm or larger. The underlying mechanism can be attributed to the distribution of 2D induced-charge densities in TMDs layers. In addition, the enhanced dipolar interaction is found to be sensitive to the free carrier screening of surrounding graphene layers, suggesting it can be efficiently tuned through dielectric engineering. Such enhanced inter-site dipolar interactions may significantly change the potential energy and the stability of the correlated insulator phase formed by dipolar and quadrupolar excitons in moiré patterns of few-layer van der Waals structures, and can also increase the importance of phonon modes as low-energy collective excitations of these correlated insulators when under external perturbations [43, 51]. Furthermore, the strong inter-site interaction can facilitate the study of novel quantum states in condensed matter physics, e.g., the charge-density wave and supersolid phases [52].

The enhanced dipolar/quadrupolar interaction strength can be detected from the photoluminescence of localized IXs in moiré patterns [21, 22]. In correlated insulators of IXs [24-28], the dipolar/quadrupolar interaction between IXs leads to blue shifts of their luminescence energies, where the interaction strength can be measured experimentally. In Ref. [53] generalized Wigner crystal states under electron fillings of $v = 1/3$ and $2/3$ have been realized in a WS$_2$/WSe$_2$ moiré pattern. When an IX is excited at a moiré potential minimum without electron occupation, it feels the dipolar repulsion from electrons localized at nearest-neighbor sites. By comparing IX emission energies between $v = 0$, 1/3 and 2/3, the dipolar interaction strength between the IX and electrons at different sites can be obtained. The experimentally measured value is found to be significantly larger than that without the 2D dielectric screening, but in decent agreement with our calculation (see the Appendix for details). Besides the luminescence energy, it is also suggested that the radiative emission rate and circular polarization of a localized IX in moiré patterns of homobilayer $MoTe_2$ depend sensitively on its center position [23]. The enhanced dipolar force can displace the IX wavepacket and substantially change its luminescence properties.

In our above analysis, the starting point is Poisson's equation for the electrostatic potential (i.e., Eq. (1)), whose validity requires the in-plane distance $r$ to be much larger than the monolayer lattice constant (~ 0.3 nm). This is generally valid, since in most cases we consider the dipolar interaction between two IXs located at two different moiré potential minima, which are separated by a typical distance of several nm. Meanwhile, the IX Bohr radius (spatial extension of the center-of-mass motion) is around 2 nm (1 nm) in realistic TMDs systems [35-39], resulting in electron and hole distribution widths much larger than the lattice constant. Eq. (1) also assumes that the out-of-plane component of the electric field introduces a negligible polarization to the few-layer van der Waals structure. However, the small but finite interlayer hopping between neighboring layers [3] can result in a finite polarization proportional to the out-of-plane electric field. A multilayer structure thus can be viewed as a dielectric slab with a finite thickness $L$ along the out-of-plane direction and anisotropic dielectric constants. We denote $\epsilon_\parallel$ ($\epsilon_\perp$) as the dielectric constant along the in-plane (out-of-plane) direction. Such a dielectric slab model has been adopted in early works [54], and it is found that in the long wavelength limit $L\sqrt{\epsilon_\parallel/\epsilon_\perp} \ll r$ the resultant electrostatic potential agrees with the 2D result obtained from Eq. (1). Note that although Eq. (1) becomes inapplicable for layers with thickness $L \sim r\sqrt{\epsilon_\perp/\epsilon_\parallel}$ or larger, the enhancement to the dipolar interaction persists as long as $\epsilon_\parallel > \epsilon_\perp$ (first-principles calculations give $\epsilon_\parallel \approx 2\epsilon_\perp \approx 16$ in TMDs [55]). In the bulk limit with $L \to \infty$, the dipolar interaction form becomes $V_D^{(aniso)}(r) \equiv \frac{2}{\sqrt{\epsilon_\perp \epsilon_\parallel}}\left(\frac{1}{r} - \frac{1}{\sqrt{r^2+D^2\epsilon_\parallel/\epsilon_\perp}}\right)$. Under a fixed average dielectric constant $\sqrt{\epsilon_\perp \epsilon_\parallel}$, $V_D^{(aniso)}(r)$ increases with $\epsilon_\parallel/\epsilon_\perp$ which is indeed enhanced compared to $V_D^{(3D)}(r) \equiv \frac{2}{\sqrt{\epsilon_\perp \epsilon_\parallel}}\left(\frac{1}{r} - \frac{1}{\sqrt{r^2+D^2}}\right)$.

**Acknowledgement.** H.Y. acknowledges support by NSFC under grant No. 12274477, and the Department of Science and Technology of Guangdong Province in China (2019QN01X061).

# Appendix. The effect of surrounding hBN layers on the dipolar interaction in TMDs

The interlayer separation $D = 1.0 + 0.34(n − 1)$ nm between the top and bottom TMDs layers can be adjusted by varying the number $n$ of hBN layers in-between. Meanwhile, the finite 2D dielectric screening of hBN will affect the dipolar interaction. Eq. (3) indicates that, the dielectric screening of hBN doesn't affect the dipolar interaction strength when $n = 1$. This is because the electric potentials from the electron and hole cancel exactly in the middle plane between the two TMDs layers thus there's no induced-charge in hBN. On the other hand, when two or more layers of hBN are inserted, induced-charges will emerge in hBN layers and the resultant dipolar interaction varies with the 2D screening length $r_0'$ of monolayer hBN (see Fig. A1(a)). Fig. A1(b) shows the comparison between $r_0' = 0$, 0.63 nm (the realistic value of monolayer hBN) and 5 nm for $n = 2$, 3 and 4. As can be seen, corrections to the dipolar interaction by $r_0'$ are extremely small. This indicates that the insertion of few-layer hBN mainly serves to adjust $D$, with a negligible dielectric screening effect on the dipolar interaction.

In experiments, TMDs layers are often encapsulated by multilayers of hBN, see Fig. A1(c) for an illustration. We assume that the top and bottom hBN layers have the same layer number $n$, whose value should affect the dipolar interaction between IXs in TMDs. The solid curves in Fig. A1(d) correspond to our calculated dipolar interaction strengths for $r_0 = 4$ nm and $n = 0$, 1, 2 and $\infty$. Here to simplify the calculation, the bilayer hBN is approximated by a single 2D layer with a screening length **2**$r_0' = 1.26$ nm and a vertical distance 0.665 nm away from TMDs, whereas the bulk hBN with $n = \infty$ is treated as a semi-infinite 3D dielectric medium following the procedure in Ref. [41] with $\epsilon_\perp = 3$ and $\epsilon_\parallel = 5$. The obtained dipolar interaction strengths under different $n$ have rather close magnitudes, again indicating that the dielectric screening effect of surrounding hBN layers is weak. We then compare these results to values extracted from the experiment in Ref. [53], where IXs are excited at empty sites in generalized Wigner crystals under electron fillings of $v = 1/3$ and 2/3 in a WS$_2$/WSe$_2$ moiré pattern (see Fig. A1(d) inset). By comparing the IX emission energy at $v = 1/3$ ($v = 2/3$) to that at $v = 0$, we get $V_D^{(B)} \approx 3.3$ meV (4.2 meV) when $r = 8$ nm which are shown as star symbols in Fig. A1(d). These values are significantly higher than $V_D^{(3D)} \approx 1.2$ meV without considering the 2D dielectric screening of TMDs and hBN, but in decent agreement with our calculations shown in Fig. A1(d). On the other hand, if we only consider the 2D screening of TMDs but treat the multilayer hBN as a background dielectric medium with $\epsilon = 4$, then Eq. (3) gives a dipolar interaction strength $\approx 0.4$ meV which is one order of magnitude smaller than the experimental value. This implies that these surrounding hBN layers cannot be simply treated as a background dielectric medium when referring to dipolar interactions in layered materials.

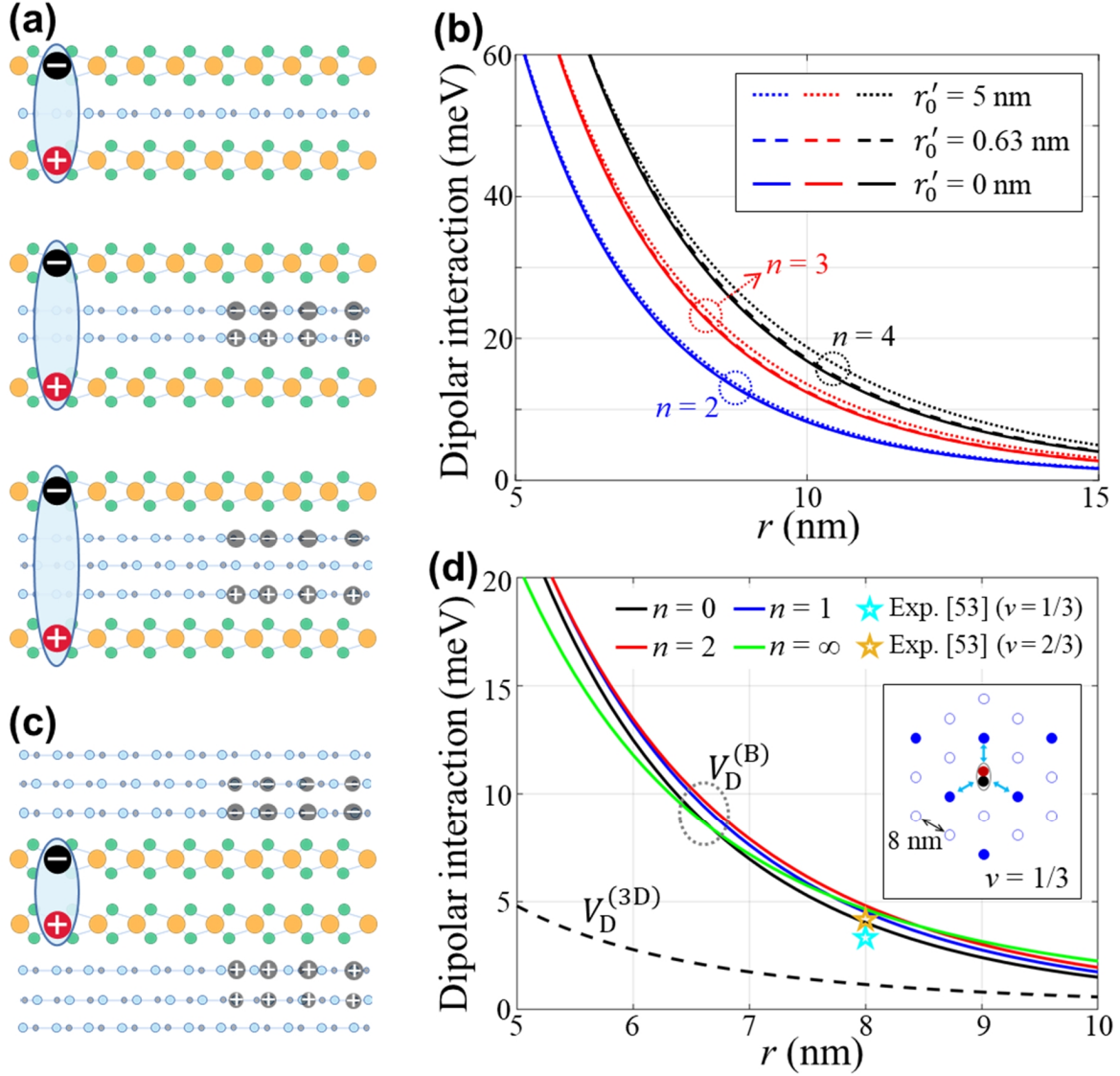

Figure A1. (a) Schematic illustrations of TMDs bilayers separated by $n = 1$, 2 and 3 layers of hBN. The induced positive and negative charges in hBN layers by an IX are shown as small grey dots with + and – symbols, respectively. (b) Dipolar interaction strengths under 2D screening length $r_0' = 0$, 0.63 and 5 nm for monolayer hBN. Blue, red and black curves are for $n = 2$ with an interlayer distance $D = 1.34$ nm, $n = 3$ with $D = 1.68$ nm, and $n = 4$ with $D = 2.02$ nm, respectively. (c) A schematic illustration of the TMDs bilayer encapsulated by two $n$-layer hBN, where small grey dots denote induced-charges in hBN by an IX in TMDs. (d) Solid curves are calculated dipolar interaction strengths for hBN layer numbers $n = 0$, 1, 2 and $\infty$, and the dashed curve is the value without considering the 2D screening of TMDs and hBN. The star symbols correspond to experimental values extracted from Ref. [53] under electron fillings $\nu = 1/3$ and 2/3. The inset illustrates the experimental scheme under $\nu = 1/3$, where solid blue dots denote localized electrons in the generalized Wigner crystal and empty dots denote sites without electron occupations. A low-energy IX excited at an empty site feels mainly the dipolar repulsion from its three nearest-neighbor electrons.